\documentclass[aps,showpacs,nofootinbib,prd,twocolumn,10pt]{revtex4-1}
\usepackage{graphicx}
\usepackage[retainorgcmds]{IEEEtrantools}
\usepackage{amssymb}
\usepackage{amsmath}
\usepackage[svgnames]{xcolor}
\usepackage{mathtools,slashed}
\usepackage{epstopdf}
\usepackage[utf8]{inputenc}
\usepackage{url}
\usepackage[colorlinks,citecolor=DarkGreen,linkcolor=DarkRed,urlcolor=DarkBlue]{hyperref}
\usepackage{subfiles}
\usepackage{physics}
\usepackage[compat=1.1.0]{tikz-feynhand}
\usepackage{tikz}
\usepackage{comment}
\usetikzlibrary{arrows.meta}
\usetikzlibrary{decorations.markings}

\usepackage{newtxtext}
\usepackage{newtxmath}
\usepackage[normalem]{ulem}

\usepackage{epsfig}

\newcommand{\be}{\begin{eqnarray}}
\newcommand{\ee}{\end{eqnarray}}
\begin{document}

\title{Mass and coupling magnetic field dependence in a scalar theory with charged bosons from an environmentally friendly renormalization group analysis}

\author{Alejandro Ayala$^{1,2,3}$}
\author{Fl\'avia Fialho$^{4}$}
\author{Ana Mizher$^{2,5}$}
\affiliation{
$^1$Instituto de Ciencias
Nucleares, Universidad Nacional Aut\'onoma de M\'exico, Apartado
Postal 70-543, CdMx 04510,
Mexico.\\
$^2$Instituto de Física, Universidade de São Paulo, Rua do Matão, 1371, CEP 05508-090, São Paulo, SP, Brazil.\\
$^3$Instituto de F\' isica Te\'orica, Universidade Estadual Paulista, Rua Dr. Bento Teobaldo Ferraz, 271 - Bloco II, 01140-070 S\~ao Paulo, SP, Brazil.\\
$^4$Laborat\'orio de F\'isica Te\'orica e Computacional, Universidade Cidade de S\~ao Paulo, 01506-000, S\~ao Paulo, Brazil\\ 
\color{black}$^5$ Centro de Ciencias Exactas and Departamento de Ciencias B\'asicas, Facultad de Ciencias, Universidad del B\'io-B\'io, Casilla 447, Chill\'an, Chile.}

\begin{abstract}

We compute the running of the mass of a neutral boson and of its self-coupling in a simple model describing the self-interaction of three scalars, one of them neutral and the other two electrically charged, subject to the effects of a magnetic field, as functions of the field strength, at one-loop order. We resort to the Environmentally Friendly Renormalization Group approach, where the flow variable is taken as that describing the environmental conditions, in this case the strength of the magnetic field. We find the magnetic field dependent mass and coupling beta functions and use them to set up the differential equations satisfied by the neutral scalar mass and coupling. We solve the resulting system of coupled equations both numerically, and also analytically in the small-mass approximation. We find that the neutral scalar mass increases, while the coupling decreases with increasing field strength. The study is intended to set up the ideas to later use them in more sophisticated theories such as QED and QCD.

\end{abstract}

\maketitle

\section{Introduction}\label{introduction}

When particles are immersed in a medium, their properties are modified. The medium can be described, for example, in terms of a statistical ensemble, such as in the case of a finite temperature or density environment. However, in general, this environment can also be provided by an external agent, such as a magnetic field. Examples of systems of particles where the effects of these fields can be important include the early universe~\cite{Grasso:2000wj}, the interior of compact astrophysical objects~\cite{Duncan:1992hi,Kouveliotou:1998ze}, and relativistic heavy-ion collisions~\cite{Kharzeev:2007jp,Skokov:2009qp,Voronyuk:2011jd}. For these systems, an interesting question to address is how to describe from a field theoretical point of view the magnetic field induced changes on the particles properties, such as their masses and couplings as functions of the magnetic field strength. For strongly interacting theories, the running of masses and couplings in either a thermal, magnetized, and thermo-magnetic medium have been studied in the context of QED, QCD and effective QCD models, see for example Refs.~\cite{Gusynin:1999pq,Farias:2014eca,Ayala:2014iba,Ayala:2016bbi,Ayala:2014uua,Mueller:2014tea,GomezDumm:2023owj,Carlomagno:2022arc,Carlomagno:2022inu,Ayala:2015bgv,Ayala:2018zat,Ayala:2020muk,Avancini:2021pmi,Coppola:2018vkw,Coppola:2023mmq,Li:2016dta,Farias:2016gmy,Ayala:2018ina,Ayala:2006sv,Steffens:2004sg,Ayala:2019akk,Ayala:2021lor,Rojas:2008sg,Castano-Yepes:2022luw,Castano-Yepes:2023brq,Castano-Yepes:2024ctr,Castano-Yepes:2024ltr,Avancini:2021pmi}.

A convenient field theoretical framework for studying the running of the masses and couplings is provided by the renormalization group (RG) approach. The essential physical observation in this context is that bare masses and couplings are fixed once and for all and are therefore independent of the energy scale -- called the subtraction point -- feeding into a given Green's function. From this observation one deduces the usual Callan-Symanzik equation for Green's functions of the theory~\cite{Callan:1970yg,Symanzik:1970rt}. This approach has been followed in Refs.~\cite{Ayala:2018wux,Ayala:2019nna} to compute the running of the QCD coupling taking as the subtraction point large values of the magnetic strength and temperature in a thermo-magnetic environment. A similar description for the running of the strong coupling with the magnetic field has been used in Refs.\cite{Fraga:2023lzn,Fraga:2023cef} (see also Ref.~\cite{Adhikari:2024bfa}) to compute the QCD pressure in a strong magnetic field in the cold-dense and hot cases. 

The RG flow of masses and couplings expresses how Green’s functions change when the arbitrary subtraction point is varied. This flow is encoded into the beta functions of the theory. However, when a system of particles is immersed in an environment, the question of interest is not necessarily how the RG flow happens when the subtraction point is varied, but instead, for a fixed subtraction point, how the RG flow happens as a function of the environment variables. This is the essence of the Environmentally Friendly Renormalization Group (EFRG)~\cite{Alexandre:1997gj,OConnor:1993sot,OConnor:1993ukd,vanEijck:1994hc}. The EFRG chooses the physical environmental parameter (temperature, density, field strength, etc.) as the RG scale, and imposes renormalization conditions at a given environmental scale. As the environment changes, the renormalized parameters run and the beta functions encode the running of these parameters with the environmental scale. 

In this work, we use the ideas of the EFRG to study the running of the coupling and the mass of a theory consisting of neutral and charged scalars with a quartic self-interaction in the presence of constant and uniform magnetic field for arbitrary values of field. The calculation is intended to gain insight and to present the method to later apply these ideas to the more sophisticated cases of QED and QCD. For this reason, we do not perform the analysis for the charged bosons mass, which we intent to perform, including the interaction between bosons and fermions in an upcoming study. The work is organized as follows: In Sec.~\ref{model} we set up the field theory of study described by a Lagrangian for a triplet of scalars, one neutral and the other two electrically charged, in the presence of a background magnetic field, and show the results of the computation of the neutral scalar two- and four-point functions up to one-loop order. The effects of the magnetic field are accounted for by considering the charged scalar propagators in the loop in terms of the Schwinger proper time representation~\cite{Schwinger:1948iu}. In Sec.~\ref{EFRG} we identify the lowest order one-loop expressions for the coupling and mass beta functions. The solutions for the coupling and the mass are given in terms of two coupled integral equations that we solve both numerically and in the approximation where the vacuum scalar mass is small, for which an analytical solution for the running with the magnetic field strength for both parameters exists. We finally present a summary and conclusions of our findings in Sec.~\ref{Concl} and leave for the appendices the explicit computation of the two- and four-point functions.

\section{Field theory of one neutral and two charged bosons in a magnetic field}\label{model}

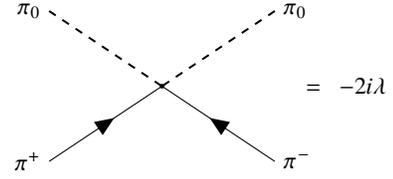
\begin{figure}[t]
\centering
\begin{tikzpicture}
\begin{feynhand}
 \vertex [dot, minimum size=1pt] (v) at (0,0) {};

  \vertex (c1) at (-1.5,1);   
  \vertex (c2) at (-1.5,-1);  
  \vertex (n1) at (1.5,1);    
  \vertex (n2) at (1.5,-1);   

  \propag[with arrow=0.5] (n2) to (v);
  \propag[with arrow=0.5] (c2) to (v);

  \propag[style=scalar, dashed] (c1) to (v);
  \propag[style=scalar, dashed] (n1) to (v);

  \node[left]  at (c1) {$\pi_0$};
  \node[left]  at (c2) {$\pi^+$};
  \node[right] at (n1) {$\pi_0$};
  \node[right] at (n2) {$\pi^-$};
\node[right] at (1.8,0) {$=\ \ -2 i \lambda$};
\end{feynhand}
\end{tikzpicture}
\caption{Feynman diagram representing the interaction Lagrangian between two neutral (dashed lines) and two charged (solid lines) scalar fields and its corresponding factor to use for the Feynman rules.}
\label{fig1}
\end{figure}

We work with a Lagrangian describing the interaction of one neutral and two charged scalars by means of a coupling $\lambda$, that in vacuum have a common mass, and where the latter are also influenced by the presence of constant and uniform magnetic field $\vec{B}=B\hat{z}$,
\begin{eqnarray}
{\mathcal{L}}&=&\left(D_\mu\pi^+\right)\left(D^\mu\pi^-\right)
+ \frac{1}{2}\left(\partial_\mu\pi_0\right)\left(\partial^\mu\pi_0\right) \nonumber\\
&-& \frac{m^2}{2}\left(2\pi^+\pi^-+\pi_0^2\right)
- \frac{\lambda}{4}\left(2\pi^+\pi^- + \pi_0^2\right)^2,
\label{Lagrangian}
\end{eqnarray}
with
\begin{eqnarray}
D_\mu\pi^\pm &=&\left(\partial_\mu \mp iqA_\mu\right)\pi^\pm,
\label{covariantD}
\end{eqnarray}
where the vector potential $A_\mu$ gives rise to the magnetic field. Notice that the interaction term can be written as
\begin{eqnarray}
{\mathcal{L}}_{\mbox{\small{int}}}&=&-\frac{\lambda}{4}\left(2\pi^+\pi^- + \pi_0^2\right)^2\nonumber\\
&=&-\frac{\lambda}{4}\left(4(\pi^+\pi^-)^2 + 4\pi^+\pi^-\pi_0^2 + \pi_0^4\right),
\label{intL}
\end{eqnarray}
from where the symmetry factors for Feynman diagram calculations can be read off. In particular, the interaction term to construct the neutral pion two- and four-point function is 
\begin{eqnarray}
{\mathcal{L}}_{\mbox{\small{int}}}^{\pi^+\pi^-\pi_0^2}=-\lambda\ \pi^+\pi^-\pi_0^2.
\end{eqnarray}
The corresponding Feynman rule to use in perturbative calculations is shown in Fig.~\ref{fig1}. The calculation of the neutral scalar self-energy, depicted in Fig.~\ref{fig2}, is shown in detail in appendix~\ref{app1}, here we just quote the result
\begin{eqnarray}
\Pi &=& \frac{\lambda(b)}{2}
A(m(b),b),\nonumber\\
A &=&  \frac{m^{2}}{4 \pi^{2}} \left[ \frac{1}{\epsilon} +\ln\left(\frac{\mu^{2}}{2b^{2}}\right)\right]   - \frac{b^{2}}{2\pi^{2}}\,\ln\left[\frac{1}{\sqrt{2\pi}\,}\,\Gamma\left(\frac{1}{2}+\frac{m^{2}}{2b^{2}}\right)\right],\nonumber\\
\end{eqnarray}
where we have defined $b^2=|qB|$.

We now turn to the computation of the four-point function for the neutral scalar. The contributing Feynman diagrams are depicted in Fig.~\ref{fig3}. The calculation is shown in detail in appendix~\ref{app2}, here we just quote the result
\begin{eqnarray}
\Lambda&=&-\frac{3}{2}\lambda^2(b)C(m(b),b)\nonumber\\
C&=&\frac{1}{4\pi^2}\left[\frac{1}{\epsilon}-\psi\left(\frac{1}{2}+\frac{m^2}{2b^2} \right) - \ln\left(\frac{2b^2}{\mu^2}\right)\right],
\end{eqnarray}
where $\psi$ is the digamma function.

\begin{figure}[b]
\centering
\begin{tikzpicture}
\begin{feynhand}

  \vertex (in)  at (-1.5,0) {};
  \vertex (v) [dot, minimum size=1pt]   at (0,0)   {}; 
  \vertex (out) at (1.5,0)   {};

  \propag[style=scalar, dashed] (in) to (v);
  \propag[style=scalar, dashed] (v)  to (out);

\draw[
    postaction={decorate},
    decoration={
        markings,
        mark=at position 0.25 with {\arrow[line width=1pt]{>}}
    }
] (0,0.6) circle (0.6);


\end{feynhand}
\end{tikzpicture}
\caption{Feynman diagram representing the neutral scalar (dashed-line) self-energy. The solid line represents a charged scalar in the loop.}
\label{fig2}
\end{figure}
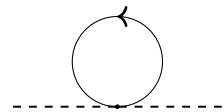
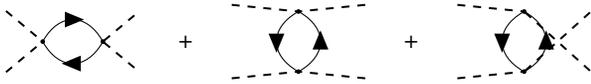
\begin{figure}[t]
\centering
\begin{tikzpicture}[scale=0.5]
\begin{feynhand}

\begin{scope}[xshift=-6cm]

  \vertex (s_a1) at (-2,  1)  {};
  \vertex (s_a2) at (-2, -1)  {};
  \vertex (s_b1) at ( 2,  1)  {};
  \vertex (s_b2) at ( 2, -1)  {};

  \vertex [dot, minimum size=1pt] (s_vL) at (-0.8, 0) {};
  \vertex [dot, minimum size=1pt] (s_vR) at ( 0.8, 0) {};

  \propag[style=scalar, dashed] (s_a1) -- (s_vL);
  \propag[style=scalar, dashed] (s_a2) -- (s_vL);
  \propag[style=scalar, dashed] (s_vR) -- (s_b1);
  \propag[style=scalar, dashed] (s_vR) -- (s_b2);

  \propag[with arrow=0.5]
    (s_vL) to [out=60,  in=120, looseness=1.4] (s_vR);
  \propag[with arrow=0.5]
    (s_vR) to [out=-120, in=-60, looseness=1.4] (s_vL);

  \node at (3,0) {$+$};

\end{scope}

\begin{scope}

  \vertex (t_aL) at (-2,  1)  {};
  \vertex (t_aR) at ( 2,  1)  {};
  \vertex (t_bL) at (-2, -1)  {};
  \vertex (t_bR) at ( 2, -1)  {};

  \vertex [dot, minimum size=1pt] (t_vT) at (0,  0.8) {};
  \vertex [dot, minimum size=1pt] (t_vB) at (0, -0.8) {};

  \propag[style=scalar, dashed] (t_aL) -- (t_vT);
  \propag[style=scalar, dashed] (t_aR) -- (t_vT);
  \propag[style=scalar, dashed] (t_bL) -- (t_vB);
  \propag[style=scalar, dashed] (t_bR) -- (t_vB);

  \propag[with arrow=0.5]
    (t_vT) to [out=-150, in=150, looseness=1.4] (t_vB);
  \propag[with arrow=0.5]
    (t_vB) to [out=  30, in=-30, looseness=1.4] (t_vT);

  \node at (3,0)  {$+$};

\end{scope}

\begin{scope}[xshift=6cm]

  \vertex (u_aL) at (-2,  1)  {};
  \vertex (u_bL) at (-2, -1)  {};
  \vertex (u_aR) at ( 2,  1)  {};
  \vertex (u_bR) at ( 2, -1)  {};

  \vertex [dot, minimum size=1pt] (u_vT) at (0,  0.8) {};
  \vertex [dot, minimum size=1pt] (u_vB) at (0, -0.8) {};

  \propag[style=scalar, dashed] (u_aL) -- (u_vT);
  \propag[style=scalar, dashed] (u_bL) -- (u_vB);
  \propag[style=scalar, dashed] (u_vT) -- (u_bR);
  \propag[style=scalar, dashed] (u_vB) -- (u_aR);

  \propag[with arrow=0.5]
    (u_vT) to [out=-150, in=150, looseness=1.4] (u_vB);
  \propag[with arrow=0.5]
    (u_vB) to [out=  30, in=-30, looseness=1.4] (u_vT);

\end{scope}

\end{feynhand}
\end{tikzpicture}
\caption{Sum of Feynman diagrams contributing to the neutral scalar four-point function. From left to right, the diagrams correspond to the $s$-, $t$- and $u$-channels, respectively. The dashed lines represent the neutral scalars whereas the solid lines represent the charged scalars in the loop.}
\label{fig3}
\end{figure}

\section{Mass and coupling running with the magnetic field strength}\label{EFRG}

Recall that the general relation between the bare-mass squared $m_B^2$ and the renormalized mass, defined at the renormalization group scale $b$ is
\begin{eqnarray}
m_B^2=m^2(b) + \delta m^2(b), 
\label{baremass}
\end{eqnarray}
where $\delta m^2(b)$ is the counterterm. The renormalization condition states that for a given value $b$
\begin{eqnarray}
\Gamma^{(2)}(p^\mu=0,m^2(b),\lambda(b),b)=m^2(b),
\end{eqnarray}
this means that at the scale $b$, the renormalized two-point function at zero momentum equals the renormalized mass $m^2(b)$. From Eq.~(\ref{Gamma2}), the one-loop two-point function is given by
\begin{eqnarray}
\Gamma^{(2)} = m^2(b) + \frac{\lambda(b)}{2}
A(m(b),b) + \delta m^2(b),
\label{Gamma2body}
\end{eqnarray}
Imposing the renormalization condition
\begin{eqnarray}
m^2(b) = m^2(b) + \frac{\lambda(b)}{2}
A(m(b),b) + \delta m^2(b),
\label{rengamma2cond}
\end{eqnarray}
we can cancel $m^2(b)$ on both sides, condition that means
\begin{eqnarray}
\delta m^2(b)= -\frac{\lambda(b)}{2}
A(m(b),b).
\end{eqnarray}
From Eq.~(\ref{baremass}), we get
\begin{eqnarray}
m_B^2=m^2(b)-\frac{\lambda(b)}{2}
A(m(b),b).
\end{eqnarray}
Since the bare mass is fixed once and for all, the renormalized mass must run with $b$ to compensate for the $b$-dependence of the loop term. This condition means that
\begin{eqnarray}
\frac{dm_B^2}{db}&=&0\nonumber\\
\frac{dm^2}{db}-\frac{1}{2}\frac{d}{db}\left[\lambda  A\right]&=&0\nonumber\\
\frac{dm^2}{db}-\frac{1}{2}\left[\frac{d\lambda}{db}A+\lambda \frac{dA}{db}\right]&=&0,
\label{massrunning}
\end{eqnarray}
where for simplicity of the notation, we have omitted the dependence on $m(b)$ and $b$ of the functions involved. Notice the following book-keeping counting in powers of $\lambda$:
\begin{itemize}
\item $A$ is a function of $m$ and $\lambda$ so it is ${\mathcal{O}}(\lambda)$,
\item $\lambda (b)$ is ${\mathcal{O}}(\lambda)$,
\item $\beta_M(b)\equiv bdm^2(b)/db$ starts at one-loop order and thus is ${\mathcal{O}}(\lambda)$,
\item The beta-function for the coupling, $\beta_\lambda$ starts at ${\mathcal{O}}(\lambda^2)$.
\end{itemize}
Therefore
\begin{itemize}
\item $\left(d\lambda (b)/db\right)A(m(b),b)$ is ${\mathcal{O}}(\lambda^2)$,
\item $\lambda (b) \left(dA(m(b),b)/db\right)$ is ${\mathcal{O}}(\lambda)$,
\item $dm^2(b)/db$ is ${\mathcal{O}}(\lambda)$.
\end{itemize}
Since we are computing $\beta_M$ to one-loop order, we can drop the ${\mathcal{O}}(\lambda^2)$ term $(d\lambda /db)A$ and thus from Eq.~(\ref{massrunning}) 
\begin{eqnarray}
\frac{dm^2(b)}{db}=\frac{\lambda (b)}{2}\frac{dA(m(b),b)}{db}.
\end{eqnarray}
From the definition of the $\beta_M$-function, we have
\begin{eqnarray}
\beta_M(b)=\frac{\lambda (b)}{2}b\frac{dA(m(b),b)}{db}.
\end{eqnarray}
Notice that this expression is still written in terms of a total derivative. However, we can use that $A(m(b),b)$ depends both explicitly and implicitly on $b$, the latter through the dependence of the running mass, namely,
\begin{eqnarray}
\frac{dA}{db}=\frac{\partial A}{\partial b} + \frac{\partial A}{\partial m^2}\frac{dm^2}{db}.
\end{eqnarray}
Notice that since we extracted a factor of $\lambda$ to write the self-energy in terms of the function $A$, this function is independent of $\lambda$ and so is $\partial A /\partial m^2$, while $dm^2/db$ is ${\mathcal{O}}(\lambda)$. This means that
\begin{eqnarray}
\frac{dA}{db}&=& \frac{\partial A}{\partial b} + {\mathcal{O}}(\lambda),\nonumber\\
\frac{dm^2}{db}&=&\frac{\lambda}{2}\left(\frac{\partial A}{\partial b} + {\mathcal{O}}(\lambda)\right).
\end{eqnarray}
Therefore, to lowest order in $\lambda$
\begin{eqnarray}
\frac{dm^2}{db}=\frac{\lambda}{2}\frac{\partial A}{\partial b},
\end{eqnarray}
and therefore, the mass beta-function becomes
\begin{eqnarray}
\beta_M=\frac{\lambda}{2}b\frac{\partial A}{\partial b}.
\label{flowM}
\end{eqnarray}

\begin{figure}[t]
    \centering
    \includegraphics[width=1\linewidth]{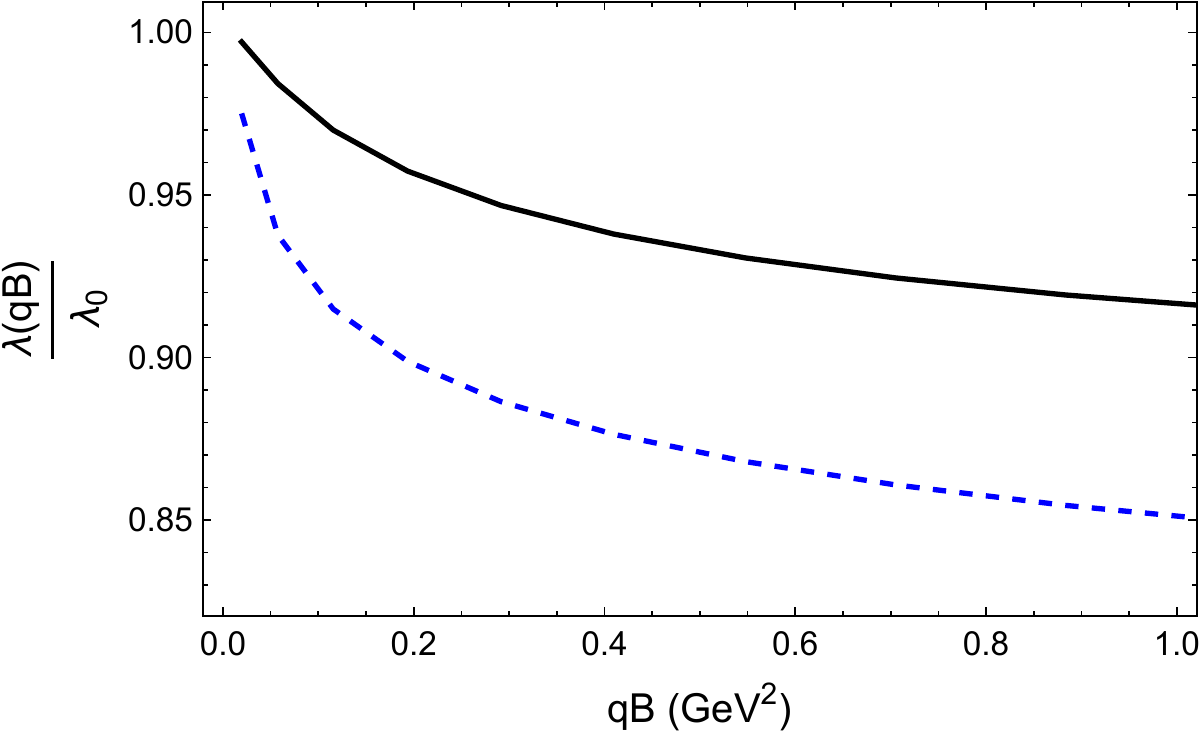}
    \caption{Magnetic field dependence of the boson self-coupling, normalized to $\lambda_0$. The solid curve represents the numerical solution, while the dashed curve represents the approximate solution. For the calculation we have taken $\lambda_0=\lambda(b_0=m_\pi)=1$ and $m_0=m(b_0)=m_\pi$, with $m_\pi=0.14$ GeV.}
    \label{fig4}
\end{figure}

\begin{figure}[t]
    \centering
    \includegraphics[width=1\linewidth]{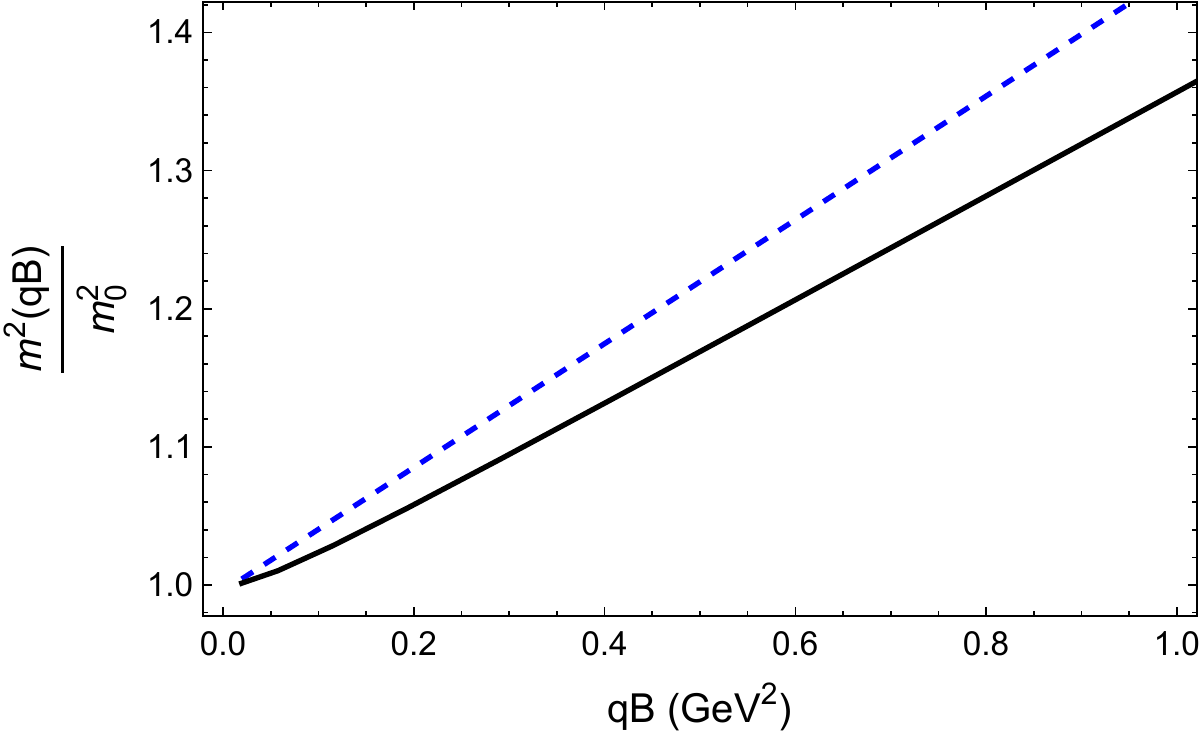}
    \caption{Magnetic field dependence of the neutral boson mass squared, normalized to $m_0^2$. The solid curve represents the numerical solution, while the dashed curve represents the approximate solution. For the calculation we have taken $\lambda_0=\lambda(b_0=m_\pi)=1$ and $m_0=m(b_0)=m_\pi$, with $m_\pi=0.14$ GeV.}
    \label{fig5}
\end{figure}

We now proceed in the same way for the coupling beta-function. At one-loop, according to Eq.~(\ref{renfourpf}), the renormalized four-point function looks like
\begin{eqnarray}
\!\!\Gamma^{(4)}(0,m^2(b),\lambda(b),b)\!=\!\lambda(b)\!-\!\frac{3}{2}\lambda^2(b)C(m(b),b) \!+\! \delta\lambda,
\end{eqnarray}
where $\delta\lambda$ is the coupling counterterm. The renormalization condition is
\begin{eqnarray}
\lambda(b)&=&\lambda(b) - \frac{3}{2}\lambda^2(b)C(m(b),b) + \delta\lambda\nonumber\\
\delta\lambda&=&\frac{3}{2}\lambda^2(b)C(m(b),b).
\end{eqnarray}
We write the relation between the renormalized and bare coupling as
\begin{eqnarray}
\lambda_B&=&\lambda(b) + \delta\lambda(b)\nonumber\\
\lambda_B&=&\lambda(b) + \frac{3}{2}\lambda^2(b)C(m(b),b).
\end{eqnarray}
Since the bare coupling does not run, we have
\begin{eqnarray}
\frac{d\lambda_B}{db}&=&0\nonumber\\
\frac{d\lambda}{db} + \frac{3}{2}\frac{d}{db}\left[\lambda^2C\right]&=&0\nonumber\\
\frac{d\lambda}{db} + \frac{3}{2}\left[2\lambda\frac{d\lambda}{db}C+\lambda^2\frac{dC}{db}\right]&=&0,
\label{runninglambda}
\end{eqnarray}
where for simplicity of notation, we have omitted the dependence on $m(b)$ and $b$ of the functions involved. Recall that
\begin{itemize}
\item $\lambda(b)$ is ${\mathcal{O}}(\lambda)$,
\item $\beta_\lambda$ is ${\mathcal{O}}(\lambda)^2$,
\item $C$ is the one-loop integral and, as a function of $\lambda$, it is ${\mathcal{O}}(0)$.
\end{itemize}
Therefore
\begin{itemize}
\item $d\lambda/db$ is ${\mathcal{O}}(\lambda^2)$,
\item $2\lambda (d\lambda /db) C$ is ${\mathcal{O}}(\lambda^3)$,
\item $\lambda^2(dC/db)$ is ${\mathcal{O}}(\lambda^2)$.
\end{itemize}
Keeping terms up to ${\mathcal{O}}(\lambda^2)$, Eq.~(\ref{runninglambda}) becomes
\begin{eqnarray}
\frac{d\lambda}{db} + \frac{3}{2}\lambda^2\frac{dC}{db}=0.
\end{eqnarray}
Recall that 
\begin{eqnarray}
\beta_\lambda\equiv b\frac{d\lambda}{db},
\end{eqnarray}
therefore
\begin{eqnarray}
\beta_\lambda=-\frac{3}{2}\lambda^2(b)b\frac{dC}{db}.
\end{eqnarray}
The function $C(m(b),b)$ depends on $b$ explicitly and also implicitly through its dependence on the running mass, therefore
\begin{eqnarray}
\frac{dC}{db}=\frac{\partial C}{\partial b} + \frac{\partial C}{\partial m^2}\frac{dm^2}{db}.
\end{eqnarray}
Notice that since we extracted a factor $\lambda^2$ to write the vertex correction in terms of the function $C$, this function is independent of $\lambda$ and so are $\partial C/\partial b$ and $\partial C/\partial m^2$, whereas $\partial m^2/\partial b$ is ${\mathcal{O}}(\lambda)$. Thus, 
\begin{eqnarray}
\frac{dC}{db}&=&\frac{\partial C}{\partial b} + {\mathcal{O}}(\lambda),\nonumber\\
\frac{d\lambda}{db}&=&-\frac{3}{2}\lambda^2\left(\frac{\partial C}{\partial b}+ {\mathcal{O}}(\lambda)\right),
\end{eqnarray}
which means that
\begin{eqnarray}
\beta_\lambda(b)=-\frac{3}{2}\lambda^2(b) b \frac{\partial C}{\partial b}.
\label{flowlambda}
\end{eqnarray}
We thus find that the flow equations are
\begin{eqnarray}
\frac{dm^2}{db}&=&\frac{\lambda}{2}\frac{\partial A}{\partial b}(m(b),b),\nonumber\\
\frac{d\lambda}{db}&=&-\frac{3}{2}\lambda^2\frac{\partial C}{\partial b}(m(b),b).
\label{flowequations}
\end{eqnarray}
Equations~(\ref{flowequations}) consist of a system of two coupled first-order ordinary differential equations. To solve it, we need to choose a reference value of the magnetic field $b_0$. At this scale we fix
\begin{eqnarray}
m^2(b_0)&=&m_0^2,\nonumber\\
\lambda(b_0)&=&\lambda_0,
\end{eqnarray}
which are chosen to reproduce a given physical input. Let us see how to formally go about solving the system. First, integrate the second of Eqs.~(\ref{flowequations}) from $b_0$ to some $b$
\begin{eqnarray}
\int_{\lambda_0}^{\lambda(b)}\frac{d\lambda'}{{\lambda'}^2}&=&-\frac{3}{2}\int_{b_0}^bdb'\frac{\partial C}{\partial b'}(m(b'),b'),\nonumber\\
\frac{1}{\lambda (b)}&=&\frac{1}{\lambda_0} + \frac{3}{2}\int_{b_0}^bdb'\frac{\partial C}{\partial b'}(m(b'),b'),\nonumber\\
\lambda(b)&=&\frac{\lambda_0}{1+\frac{3}{2}\lambda_0\int_{b_0}^bdb'\frac{\partial C}{\partial b'}(m(b'),b')}.
\label{lrunning}
\end{eqnarray}
Now let us look at the first of Eqs.~(\ref{flowequations}) integrating from $b_0$ to some $b$
\begin{eqnarray}
m^2(b)=m_0^2 + \frac{1}{2}\int_{b_0}^b
db'\lambda(b')\frac{\partial A}{\partial b'}(m(b'),b').
\label{mrunning}
\end{eqnarray}
Given the explicit expressions for $A$ and $C$, the problem reduces to finding the solutions for the system of equations that relate $m^2(b)$ and $\lambda(b)$. In order to get intuition, before presenting the numerical solution, we first work with the small mass approximation for which 
\begin{eqnarray}
A(m(b),b)&=&A(0,b)\nonumber\\
C(m(b),b)&=&C(0,b).
\end{eqnarray}
ignoring the (magnetic field independent) divergent part, which is taken care by the counterterm
\begin{eqnarray}
A(0,b)&\to&\frac{b^2}{4\pi^2}\ln(2)\nonumber\\\
C(0,b)&\to&\frac{1}{4\pi^2}\left[\psi\left(\frac{1}{2}\right) + \ln\left(\frac{2b^2}{\mu^2}\right)\right],
\end{eqnarray}
which implies that
\begin{eqnarray}
\frac{\partial A}{\partial b}&=&\frac{b}{2\pi^2}\ln(2)\nonumber\\
\frac{\partial C}{\partial b}&=&\frac{1}{2\pi^2 b}.
\end{eqnarray}
Plugging these expressions into Eqs~(\ref{lrunning}) and~(\ref{mrunning}), we obtain
\begin{eqnarray}
\lambda(b)&=&\frac{\lambda_0}{1+\frac{3}{4\pi^2}\lambda_0\int_{b_0}^b\frac{db'}{b'}}\nonumber\\
\lambda(b)&=&\frac{\lambda_0}{1+\frac{3}{4\pi^2}\lambda_0\ln\left(\frac{b}{b_0}\right)},
\label{lambdaapprox}
\end{eqnarray}
and
\begin{eqnarray}
m^2(b)=m_0^2 + \frac{\ln(2)}{4\pi^2}\lambda_0\int_{b_0}^b db'\frac{b'}{1+\frac{3}{4\pi^2}\lambda_0\ln\left(\frac{b'}{b_0}\right)}.
\label{mapprox0}
\end{eqnarray}
The integral in Eq.~(\ref{mapprox0}) can be found analytically, albeit in terms of transcendental functions
\begin{eqnarray}
\int_{b_0}^b db'\frac{b'}{1+\frac{3}{4\pi^2}\lambda_0\ln\left(\frac{b'}{b_0}\right)}&\!\!=\!\!&-
\frac{4\pi^2}{3\lambda_0}b_0^2e^{-\frac{8\pi^2}{3\lambda_0}}\nonumber\\
&\!\!\times\!\!&\left[{\mbox{ExpIntEi}}\left(\frac{8\pi^2}{3\lambda_0}\right)\right.\nonumber\\
&\!\!-\!\!&\left.{\mbox{ExpIntEi}}\left(\frac{8\pi^2}{3\lambda_0}+\ln\left(\frac{b^2}{b_0^2}\right)\right)\right],\nonumber\\
\end{eqnarray}
where ExpIntEi is the exponential integral function Ei. Expanding up to ${\mathcal{O}}(\lambda)$ we get
\begin{eqnarray}
\int_{b_0}^b db'\frac{b'}{1+\frac{3}{4\pi^2}\lambda_0\ln\left(\frac{b'}{b_0}\right)}&\!\!\to\!\!&\frac{1}{2}(b^2-b_0^2)+\frac{3}{16\pi^2}\lambda_0\nonumber\\
&\!\!\times\!\!&\left[(b^2-b_0^2) + 2b^2\ln\left(\frac{b_0}{b}\right)\right].\nonumber\\
\end{eqnarray}
Therefore, keeping terms up to ${\mathcal{O}}(\lambda)$, we get
\begin{eqnarray}
m^2(b)=m_0^2 + \frac{\ln(2)}{8\pi^2}(b^2-b_0^2)\lambda_0.
\label{mapprox}
\end{eqnarray}
The analysis shows that the coupling $\lambda(qB)$ decreases, while the mass squared $m^2(qB)$ increases as the field strength $qB$ increases. The numerical solution of Eqs.~(\ref{lrunning}) and~(\ref{mrunning}), compared to the small mass approximate solutions of Eqs.~(\ref{lambdaapprox}) and~(\ref{mapprox}), are shown in Figs.~\ref{fig4} and~\ref{fig5}, respectively. For the computation, we have taken $\lambda_0=\lambda(b_0=m_\pi)=1$ and $m_0=m(b_0)=m_\pi$, with $m_\pi=0.14$ GeV.

\section{Summary and conclusions}\label{Concl}

The study of the change of particle properties under the influence of environmental effects, such as temperature, density, and external fields, has been a subject of intense scrutiny in recent times. Different approaches have been used to extract this information, but among them, only a few have resorted to studying the RG flow with the environmental scale. In this work, we have made use of the ideas of the EFRG approach, which consist of fixing the subtraction scale (the momentum feeding into a given Green's function) and using the scale associated with the environment as the flow variable. We have explored such an approach in a simple field theoretical model describing the self-interaction of three scalars, one of them being neutral and the other two being electrically charged, in the presence of a magnetic field, to study the running of the neutral scalar mass and coupling with the field strength at one-loop order. We have computed the magnetic field dependent mass and coupling beta functions and used them to set up the differential equations satisfied by the neutral scalar mass and coupling. The equations turn out to form a set of coupled first-order ordinary differential equations that we have solved both numerically and analytically in the small-mass approximation. We have found that the coupling decreases while the neutral boson mass increases as the field strength increases. 

The ideas discussed in this work serve as a test ground to later be used in more sophisticated theories, such as QED and QCD. Perhaps even at a more immediate reach is the study of how the evolution of the running of the neutral scalar and the coupling changes when the analysis includes the evolution of the charged scalars mass and the theory includes fermions. These ideas are currently being pursued and will be reported soon elsewhere.

\section*{Acknowledgments}

 A.A. thanks the colleagues and staff of Universidade de São Paulo, of Instituto de F\'isica Te\'orica, UNESP and of Universidade Cidade de São Paulo for their kind hospitality during a sabbatical stay. A.A. acknowledges support from the PASPA program of DGAPA-UNAM for the sabbatical stay during which this research was carried out. Support for this work has been received in part by a SECIHTI-M\'exico grant number CIORGANISMOS-2025-17 and by the DGAPA-PAPIIT-UNAM grant number IG100826. This study was financed, in part, by the São Paulo Research Foundation (FAPESP), Brazil. Process Numbers 2023/08826-7 and 2024/18493-8. F.F. acknowledges support from the Coordena\c{c}\~ao de Aperfei\c{c}oamento de Pessoal de N\'\i vel Superior (CAPES) - Finance Code 001.

\appendix
\section{Two-point function}\label{app1}

Since the one-loop contribution to the neutral scalar self-energy is momentum-independent, we can perform the calculation directly in Euclidean space where the relation between the dressed propagator $\Delta$, the free propagator $\Delta_B$ and the self-energy $\Pi$ is given by
\begin{eqnarray}
\Delta^{-1}=\Delta_B^{-1}+\Pi,
\end{eqnarray}
where
\begin{eqnarray}
\Delta_B=-\int_0^\infty\frac{ds}{\cosh(qBs)}e^{-s\left(\omega_n^2+k_3^2+k_\perp^2\frac{\tanh(qB)}{qBs}+m^2\right)},
\label{scalarpropEuc}
\end{eqnarray}
where $\omega_n$ are boson Matsubara frequencies, $B$ is the external magnetic field, $\vec{k}=(\vec{k}_\perp,k_3)$ the loop momentum, and $m$ and $q$, the boson mass and electric charge, respectively. The expression for the one-loop self-energy is given by
\begin{eqnarray}
\Pi=2\lambda T\sum_n\int\frac{d^3k}{(2\pi)^3}\Delta_B.
\end{eqnarray}
Since the loop starts and ends at the same point, the Schwinger phase vanishes. Performing the sum over Matsubara frequencies and keeping only the purely magnetic contribution, one gets
\begin{equation}
    \Pi = -2\lambda\left(\frac{|qB|}{4\pi}\right) \sum^{\infty}_{l=o}  \int \frac{dk_{3}}{2\pi}  \frac{1}{\sqrt{k_{3}^{2} + (2l+1)|qB| +m^{2}}}.
\end{equation}
This expression is logarithmically divergent. To isolate the divergence, we resort to using dimensional regularization. The dimension is shifted from $d = 1$ to $d=1-2\epsilon$, and a parameter $\mu$, with energy dimensions, is included to keep the overall expression dimensionally correct. It then follows that
\begin{eqnarray}
    \Pi &=& -2\lambda\left(\frac{|qB|}{4\pi}\right)   (\mu)^{1-d}\nonumber\\ 
    &\times&\sum^{\infty}_{l=o}  \int \frac{d^{d}k_{3}}{(2\pi)^{d}}  \frac{1}{\sqrt{k_{3}^{2} + (2l+1)|qB| +m^{2}}}.
\end{eqnarray}
Performing the integration over $k_{3}$ for which we use the expression
\begin{equation}
     \int \frac{d^{d}l}{(2\pi)^{d}} \, \frac{1}{(l^{2}+\Delta)^{n}} = \frac{(-1)^{n}}{(4\pi)^{d/2}} \,\frac{\Gamma(n-\frac{d}{2})}{\Gamma(n)} \, \left(\frac{1}{\Delta}\right)^{n-\frac{d}{2}},
\end{equation}
where $\Gamma(x)$ is the gamma function, we obtain
\begin{equation}
\label{pi.}
    \Pi = -2\lambda\left(\frac{|qB|}{8\pi^{2}}\right)(4\pi\mu^{2})^{\epsilon} \,\Gamma(\epsilon) \sum^{\infty}_{l=0} \frac{1}{[(2l+1)|qB| + m]^{\epsilon}}.
\end{equation}
In Eq.~(\ref{pi.}) we recognize the Riemann-Hurwitz Zeta function:
\begin{equation}
    \zeta(s,a) = \sum^{\infty}_{k=0} \,\frac{1}{(k+a)^{s}}; 
\end{equation}
so the self-energy becomes:
\begin{equation}
    \Pi = -\lambda\frac{|qB|}{4\pi^{2}}\, \left(\frac{4\pi\mu^{2}}{|qB|}\right)^{\epsilon}\, \,\Gamma(\epsilon)\,\,\zeta\left(\epsilon,\, \frac{1}{2}+\frac{m^{2}}{2|qB|}\right).
\end{equation}
We expand each function separately around $\epsilon\rightarrow 0$. First, the gamma function reads
\begin{equation}
\label{expgamma}
    \Gamma(\epsilon) \approx \frac{1}{\epsilon} - \gamma_{E};
\end{equation}
where $\gamma_{E}$ is the Euler-Mascheroni constant. Second, the  Riemann-Hurwitz Zeta function becomes
\begin{equation}
   \begin{split}
    \zeta\left(\epsilon, \frac{1}{2}+\frac{m^{2}}{2|qB|}\right) &\approx \zeta\left(0,\frac{1}{2}+\frac{m^{2}}{2|qB|}\right) \\ &+\epsilon \,\ln\left[\frac{1}{\sqrt{2}}\Gamma\left(\frac{1}{2}+\frac{m^{2}}{2|qB|}\right)\right],
    \end{split}
\end{equation}
and finally the factor with the $\epsilon$ in the exponent becomes
\begin{equation}
    \left(\frac{4\pi\mu^{2}}{|qB|}\right)^{\epsilon} \approx 1+\epsilon\ln\left(\frac{4\pi\mu^{2}}{|qB|}\right).
\end{equation}

Taking the limit $\epsilon\rightarrow 0$ and using the $\overline{\text{MS}}$ regularization scheme where $\mu^{2} \rightarrow \mu^{2}\,e^{\gamma_{E}}/4\pi$; the self-energy can be written as:
\begin{equation}
\begin{split}
   \Pi &= -\lim_{\epsilon\rightarrow 0} \,\lambda \frac{|qB|}{4\pi^{2}} \bigg\{\zeta\left(0, \,\frac{1}{2} +\frac{m^{2}}{2|qB|} \right) \times\\&\left[\frac{1}{\epsilon}\,+ \,\ln\left(\frac{\mu^{2}}{2|qB|}\right)\right] + \,\ln\left[\frac{1}{\sqrt{2\pi}\,}\,\Gamma\left(\frac{1}{2}+\frac{m^{2}}{2|qB|}\right)\right]\bigg\}.
    \end{split}
\end{equation}
Notice that when $s=0$, the Riemann-Hurwitz Zeta function is simply given by
\begin{equation}
    \zeta\left(0, \,\frac{1}{2} +\frac{m^{2}}{2|qB|}\right) =  - \frac{m^{2}}{2|qB|};
\end{equation}
hence, we have the following expression for the self-energy:
\begin{equation}
\begin{split}
    \Pi &= \lim_{\epsilon\rightarrow0} \,\lambda\frac{|qB|}{4\pi^{2}} \bigg\{\frac{m^{2}}{2|qB|}\left[\frac{1}{\epsilon}\,+ \,\ln\left(\frac{\mu^{2}}{2|qB|}\right)\right] \\ &- \,\ln\left[\frac{1}{\sqrt{2\pi}\,}\,\Gamma\left(\frac{1}{2}+\frac{m^{2}}{2|qB|}\right)\right]\bigg\},
    \end{split}
\end{equation}
which shows that the UV divergence is magnetic field-independent.

The self-energy contributes to the two-point function $\Gamma^{(2)}$. We choose the subtraction point as $p^\mu=0$, where $p^\mu$ is the neutral scalar four-momentum. Therefore, we have
\begin{eqnarray}
\Gamma^{(2)} = m^2(b) + \frac{\lambda(b)}{2}
A(m(b),b) + \delta m^2(b),
\label{Gamma2}
\end{eqnarray}
where we have introduced the notation $b^2=|qB|$ and have factored out the coupling from the expression of the self-energy defining 
\begin{equation}
    A =  \frac{m^{2}}{4 \pi^{2}} \left[ \frac{1}{\epsilon} +\ln\left(\frac{\mu^{2}}{2b^{2}}\right)\right]   - \frac{b^{2}}{2\pi^{2}}\,\ln\left[\frac{1}{\sqrt{2\pi}\,}\,\Gamma\left(\frac{1}{2}+\frac{m^{2}}{2b^{2}}\right)\right].
\end{equation}
We now impose the EFRG renormalization condition such that at a given value $b$, the two-point function coincides with a known value of $m^2(b)$, namely
\begin{eqnarray}
m^2(b) = m^2(b) + \frac{\lambda(b)}{2}
A(m(b),b) + \delta m^2(b).
\end{eqnarray}
Canceling $m(b)$ from the left- and right-hand sides of the above equation, we obtain the explicit expression for the counterterm
\begin{equation}
\delta m^2(b) = -\frac{\lambda(b)}{2}A(m(b),b).
\end{equation}

\section{Four-point function}\label{app2}

Considering the case where $p^\mu=0$, the contribution from the three crossed channels, depicted in Fig.~\ref{fig3} is the same, therefore, the vertex correction at one-loop level is given by
\begin{equation}
   \begin{split}
    \Lambda (0;qB) &= \frac{3(2i\lambda)^{2}}{2!} \int\frac{d^{4}k}{(2\pi)^{4}}\, \Delta_{B}(k)\, \Delta_{B}(k),
    \end{split}
    \label{fourpointexpression}
\end{equation}
where $\Delta_{B}$ is given by Eq.~(\ref{scalarpropEuc}). Notice that Eq.~(\ref{fourpointexpression}) is logarithmically UV divergent.
Also, since the particles circling the loop have the same charge, the Schwinger phase vanishes. The momentum components are denoted by $k_{\parallel}^{2} = k_{0}^{2}+k_{3}^{2}$ and $k_{\perp}^{2}=k_{1}^{2}+k_{2}^{2}$. Combining the exponentials from both propagators, we obtain
\begin{equation}
    \begin{split}
        & e^{-s_{1}\left(k_{\parallel}^{2}+k_{\perp}^{2}\frac{\tanh(qBs_{1})}{qBs_{1}}+m^{2}\right)} e^{-s_{2}\left(k_{\parallel}^{2}+k_{\perp}^{2}\frac{\tanh(qBs_{2})}{qBs_{2}}+m^{2}\right)} \\ 
        & = e^{-(s_{1}+s_{2})(k_{\parallel}^{2}+m^{2})}e^{-\frac{k_{\perp}^{2}}{qB}\frac{\sinh(qB(s_{1}+s_{2}))}{\cosh(qBs_{1})\cosh(qBs_{2})}}.
    \end{split}
\end{equation}
We can perform a Gaussian integration over $k_{\perp}$, and find
\begin{equation}
    \begin{split}
        &\Lambda(0;qB) = -\frac{3qB\lambda^{2}}{8\pi^{2}}\int^{\infty}_{0}ds_{1}\int^{\infty}_{0}ds_{2}  \\ & \times \int \frac{d^{2}k_{\parallel}}{(2\pi)^{2}} \frac{e^{-(s_{1}+s_{2})(k_{\parallel}^{2}+m^{2})}}{\sinh(qB(s_{1}+s_{2}))}.
    \end{split}
\end{equation}
Once again, we perform a Gaussian integration, this time over $k_{\parallel}$ and obtain
\begin{equation}
\begin{split}
    &\Lambda(0;qB) = -\frac{3qB\lambda^{2}}{8\pi^{2}} \\ & \times \int^{\infty}_{0}ds_{1}\int^{\infty}_{0}ds_{2} \frac{e^{-(s_{1}+s_{2})m^{2}}}{(s_{1}+s_{2})\sinh(qB(s_{1}+s_{2}))}.
    \end{split}
\end{equation}
Now, to integrate over $s_{1}$ and $s_{2}$ we make a change of variables $s=s_{1}+s_{2}$ and $u=s_{1}/(s_{1}+s_{2})$. The Jacobian of the transformation is given by $s$. Therefore, we can write:
\begin{equation}
    \Lambda(0;qB) =  -\frac{3qB\lambda^{2}}{8\pi^{2}} \int^{\infty}_{0}ds \int^{1}_{0} du\frac{e^{-sm^{2}}}{\sinh(qBs)}.
\end{equation}
The $u$ integral just gives 1. Now, to evaluate the $s$ integral, notice there is a divergence when $s=0$, so to isolate it, we regulate it with the dimension of $s$, as
\begin{equation}
\label{regulated}
    \int^{\infty}_{0}ds \frac{e^{-sm^{2}}}{\sinh(qBs)} \rightarrow \mu^{2\epsilon} \int^{\infty}_{0} ds \frac{s^{\epsilon}e^{-sm^{2}}}{\sinh(qBs)},
\end{equation}
where $\mu$ has dimensions of energy. The integral in Eq.~(\ref{regulated}) yields
\begin{equation}
    \begin{split}
    &\mu^{2\epsilon} \int^{\infty}_{0} ds \frac{s^{\epsilon}e^{-sm^{2}}}{\sinh(qBs)} \\& = \mu^{2\epsilon}2^{-\epsilon}(qB)^{-1-\epsilon}\Gamma(1+\epsilon)\zeta\left(1+\epsilon, \frac{qB+m^{2}}{2qB}\right).
    \end{split}
\end{equation}
Now, we can expand each function separately around $\epsilon\rightarrow 0$. First, the Riemann-Hurwitz Zeta reads
\begin{equation}
    \zeta\left(1+\epsilon, \frac{qB+m^{2}}{2qB}\right) = \frac{1}{\epsilon} - \psi\left(\frac{qB+m^{2}}{2qB}\right),
\end{equation}
where $\psi(a)$ is the digamma function. Second, because of the relation $\Gamma(\epsilon+1) = \epsilon\Gamma(\epsilon)$, we can use Eq.~(\ref{expgamma}) for the $\Gamma$ function. The factor with $\epsilon$ in the exponential is obtained using the general expression $a^{-\epsilon} \approx 1-\epsilon\ln(a)$. Therefore, putting all together, the expression for the one-loop vertex correction reads as
\begin{equation}
    \Lambda(0;b) = -\frac{3\lambda^{2}}{8\pi^{2}}\left[\frac{1}{\epsilon}-\psi\left(\frac{b^{2}+m^{2}}{2b^{2}
    }\right)-\ln\left(\frac{2b^{2}}{\mu^{2} }\right)\right].
\end{equation}
Notice that also for the case of the vertex correction, the UV divergence is magnetic field-independent. The four-point function, with the three-level and coupling counterterm, can be written as
\begin{equation}
    \Gamma^{(4)} = \lambda(b) - \frac{3}{2}\lambda^{2}(b)C(m(b),b)+\delta\lambda(b),
    \label{renfourpf}
\end{equation}
with $C$ given by
\begin{equation}
    C = -\frac{1}{4\pi^{2}}\left[\frac{1}{\epsilon}-\psi\left(\frac{b^{2}+m^{2}}{2b^{2}
    }\right)-\ln\left(\frac{2b^{2}}{\mu^{2} }\right)\right].
\end{equation}
Imposing the renormalization condition
\begin{equation}
    \lambda(b) = \lambda(b)-\frac{3}{2}\lambda^{2}(b)C(m(b),b)+\delta\lambda(b)
\end{equation}
we finally obtain that the counterterm is given by
\begin{equation}
    \delta \lambda(b) = \frac{3}{2}\lambda^{2}(b)C(m(b),b).
\end{equation}

\bibliography{biblio}{}
\bibliographystyle{apsrev4-1}

\end{document}